\documentclass{elsart}
\usepackage{natbib}
\usepackage{amssymb}

\def\b{\begin{equation}}
\def\e{\end{equation}}

\begin{document}
\begin{frontmatter}
\title{Why the observed Black Hole Candidates do not show spin pulsation}
\author{Abhas Mitra}
\address{ Theoretical Astrophysics Section, BARC, Mumbai-400085, India, Abhas.Mitra@mpi-hd.mpg.de, amitra@barc.ernet.in}
\begin{abstract}
It is shown here that the so-called spinning Black Hole Candidates (BHCs)
 cannot be Black Holes (BHs) at all and, on the other hand,
 must be spinning hot  Eternally Collapsing Objects (ECOs). The question would then
arise why the spinning BHCs do not reveal their spin pulsation
unlike spinning cold Neutron Stars (NSs)\citep{narayan1}. Even for
magnetized NSs with ``hard surfaces'', there may be no observed spin
pulsation if the dipole axis and spin direction are exactly same.
However, spinning ECOs do not show spin pulsation because their
surface gravitational red shift is extremely high, $z \gg 1$ and
could be in the range of $ 10^{7-8}$ for stellar mass range. In
contrast NSs have $z \sim 0.1 -0.3$. The time profile of
 periodic signals generated
on the surface of a spinning MECO get extremely distorted due to
extreme general relativistic frame dragging effect as they traverse
through extremely steep gravitational field. However, if any
radiation would be formed in the ECO magnetosphere sufficiently away
from the surface  like in the pulsar ``outer slots''(which may
happen for isolated spinning ECOs), it might be a pulsed one.
\end{abstract}
\end{frontmatter}
\section{Introduction}
When a  self-gravitating fluid undergoes gravitational contraction, by virtue
of Virial Theorem,
part of the self-gravitational energy
must be radiated out. Thus  the total mass energy, $M$,
($c=1$) of a body decreases as its  radius $R$ decreases.  But in Newtonian regime ($2M/R\ll 1$, $G=1$),
$M$ is almost fixed and  the evolution of the
ratio, $2 M/R $,  is practically dictated entirely by $R$.
 If it is {\em assumed} that even in the extreme general relativistic
case $2M/R $ would behave in the {\em same Newtonian} manner,  then
for sufficiently small $R$, it would be possible to have  $2M/R >
1$, i.e, trapped surfaces would form. Unfortunately, even when we
use General Relativity (GR), our intuition is often governed by
Newtonian concepts, and thus, intuitively, it appears that, as a
fluid would collapse, its gravitational mass would remain more or
less constant so that for continued collapse, sooner or later, one
would have $2 M/R >1$, i.e, a ``trapped surface'' must form. The
singularity theorems thus start with the {\em assumption} of
formation of trapped surfaces. In the following we show that,
actually, trapped surfaces do not form: The spherically symmertic
metric for an arbitrary fluid, in terms of comoving coordinates $t$
and $r$ is \citep{mit1, mit2}
\begin{equation}
ds^2 = g_{00} dt^2 + g_{rr} dr^2 - R^2 (d\theta^2 + \sin^2\theta d\phi^2)
\end{equation}
where $R=R(r, t)$ is the  circumference coordinate and happens to be
 a scalar. Further,  for radial motion with $d\theta =d\phi =0$,
the metric becomes
\begin{equation}
ds^2 = g_{00} ~dt^2 (1- x^2); \qquad (1-x^2) = {1\over g_{00}} {ds^2\over dt^2}
\end{equation}
where the auxiliary parameter $ x = {\sqrt {-g_{rr}} ~dr\over \sqrt{g_{00}}~ dt}$.
 The comoving observer at $r=r$ is
free to do measurements of not only the fluid element at $r=r$ but also of other objects:
  If the
comoving observer is compared with a static floating boat in a flowing
river, the boat can monitor the motion  the pebbles fixed
on the river bed. Here the fixed markers on the river bed are like the
background $R=  constant$ markers against which the
river flows. If we intend to find the parameter $x$ for such a $R=constant$
marker, i.e, for a pebble lying on the river bed at a a {\em fixed} $R$, we
will have,
$ d R(r,t) = 0= {\dot R} dt + R^\prime dr $,
where an overdot denotes a partial derivative w.r.t. $t$ and a prime denotes
a partial derivative w.r.t. $r$.
Therefore for the $R=constant$ marker, we find that
${dr\over dt} = - {{\dot R}\over R^\prime}$
and the corresponding $x$ is
\begin{equation}
x= x_{c} = {\sqrt {-g_{rr}} ~dr\over \sqrt{g_{00}}~ dt} = -{\sqrt {-g_{rr}}
~{\dot R}\over \sqrt{g_{00}}~ R^\prime}
 \end{equation}
Using Eq.(2), we also have, for the $R=constant$ pebble,
\begin{equation}
(1-x_c^2) = {1\over g_{00}} {ds^2\over dt^2}
\end{equation}
Now let us define
\begin{equation}
\Gamma = {R^\prime\over \sqrt {-g_{rr}}}; \qquad U = {{\dot R}\over \sqrt{g_{00}}}
\end{equation}
so that Eqs. (3)and (5)  yield $x_c = {U\over \Gamma}; \qquad U=
-x_c \Gamma$. As is well known, the gravitational mass of the
collapsing (or expanding) fluid is defined through the
equation\citep{mit2}
\begin{equation}
\Gamma^2 = 1 + U^2 - {2M(r,t)\over R}
\end{equation}
Using $U=-x_c ~\Gamma$ in this equation and then  transposing, we obtain
\begin{equation}
\Gamma^2 (1- x_c^2) = 1- {2M(r,t)\over R}
\end{equation}
By using Eqs.(4) and (5) in the foregoing Eq., we have
\begin{equation}
{{R^\prime}^2\over {-g_{rr} g_{00}}} {ds^2\over dt^2} = 1 - {2M(r,t)\over R}
\end{equation}
Recall that the determinant of the metric tensor  $g = R^4 \sin^2
\theta ~g_{00} ~g_{rr} \le 0$ so that we must always have $-g_{rr}~
g_{00} \ge 0$. But $ds^2 \ge 0$ for all material particles or
photons. Then it follows that the LHS of Eq.(8) is {\em always
positive}. So must then be the RHS of the same Eq. and which implies
that $2M(r,t)/ R \le 1$. Therefore trapped surfaces are not formed
in spherical collapse.
\section{ECO and absence of Pulsations}
  In case it would be {\em assumed} that, the collapse would continue
all the way upto $R=0$ to become a point, then the above constraint
demands that $M (point, R_0=0) =0$ too. This is exactly what was
found in 1962\citep{ADM}: ``$M\to 0 ~as~ \epsilon \to 0$,  and
``$M=0 ~for ~a~neutral~particle$.''  This is the reason that neutral
BHs (even if they would be assumed to exist) must have $M=0$.
However, mathematically, there could be charged finite mass BHs. But
since astrophysical compact objects are necessarily neutral, the the
{\em finite} mass BHCs and are not {\em zero} mass BHs. Sufficiently
massive bodies collapse beyond
   the NS stage and eventually become BH with an EH ($z=\infty$).
   As the collapse proceeds beyond the stage of $(1+z) =\sqrt{3}$,
   the emission cone of the radiation emitted by the body stars shrinking due to large gravitational bending of
   photons and neutrinos.
    At high $z$, the escape probability of emitted radiation thus decreases as $\sim
   (1+z)^{-2}$ and consequently pressure of the trapped radiation starts increasing by a factor $\sim (1+z)$.
   Much before, $z \to \infty$, to become a BH, trapped radiation pressure must halt the collapse dynamically as
   it would correspond to local Eddington value. This is the reason that  a ECO is born \citep{mit3, mit4}. It is likely that the magnetic field of the object
   gets virialized and becomes extremely strong \citep{rl1, rl2}. Even otherwise, the intrinsic magnetic  field
   of the ECO must be very high. Though the collapse  still proceeds to attain the
   $M=0$, $z=\infty$ BH stage, it can do so only asymptotically.
   The ECO surface radius $R_0 \approx 2M$.
 Since  $(1+z) =
(1-2M/R_0)^{-1/2}$,
   $z$ falls off sharply as one moves away from the surface. For instance at,  $R= 3 R_0$, $z\approx 0.2$, even though, $z \sim 10^{7-8}$! This variation in itself, would only reduce  the energy of radiation
   quanta by a factor $(1+ z)$. But when such an object with strong gravity ($z \gg 1$) spins, it drags it surrounding spacetime and the local
   inertial frames at various spatial locations rotate at decreasing rate $\sim R^{-3}$. Thus the phase of the light
   house signal gets constantly stretched  and distorted by a varying factor at various spatial locations.
   Consequently, no spin pulsation is seen by a distant observer.
However, for isolated ECOs, if there would be generation of
radiation away from the surface like in ``outer gaps'', then the
production region would be in a low $z$ region and the degree of
frame dragging would be comparable to that due to a pulsar. Such a
signal could be pulsed. Note that recently it has been shown that
the so-called supermassive BHs are actually supermassive
ECOs\citep{slr, rl3}.


\begin{thebibliography}{}
\bibitem[Arnowitt, Deser \& Misner(1962)]{ADM}  Arnowitt, R.,  Deser, S. \&  Misner, C.W.,
The Dynamic of General Relativity, in {\it Gravitation : an
Introduction to Current Research}, (ed. L. Witten, Wiley, NY, 1962),
p.227-265, (gr-qc/0405109).
\bibitem[Mitra(2004)]{mit1} Mitra, A., A New Proof for
Non-occurrence of Trapped Surfaces and Information Paradox,
astro-ph/0408323, p.1.-p6
\bibitem[Mitra(2005)]{mit2}  Mitra, A., BHs or ECOs: A Review of 90 Years of Misconceptions,
in {\it Focus on Black Holes Research}, p1- p94, (Nova Science Pub.,
NY, 2005)
\bibitem[Mitra(2006a)]{mit3} Mitra, A., A Generic Relation Between Baryonic and
Radiative Energy Densities of Stars, MNRAS Lett., 367, L66-L68, 2006
(gr-qc/0601025)
\bibitem[Mitra(2006b)]{mit4} Mitra, A., Radiation Pressure Supported Stars in
Einstein Gravity: ECOs, MNRAS (in press), 2006, (gr-qc/0603055)
\bibitem[Narayan(2005)]{narayan1}  Narayan, R., Black Holes in Astrophysics, New J. Phys, 7, 199-218, 2005



\bibitem[Robertson \& Leiter(2002)]{rl1}  Robertson, S. and  Leiter, D., Evidence for Intrinsic Magnetic Moment in Black Hole Candidates,
Astrophys. J., 565, 447-451, 2002 (astro-ph/0102381)
\bibitem[Robertson \& Leiter(2005)]{rl2}  Robertson, S. and  Leiter, D., MECO Model of Galactic Black Hole Candidates and Active Galactic Nuclei,
in ``New Developments in Black Hole Research'', p1-p44, (Nova
Science Pub., NY, 2005),  (astro-ph/0602453)
\bibitem[Robertson \& Leiter(2006)]{rl3}  Robertson, S. and  Leiter,
D., Does Sgr A* Have an Intrinsic Magnetic Moment Instead of an
Event Horizon?, Astrophysical J. (submitted, 2006), astro-ph/0603746
\bibitem[Schild, Leiter \& Robertson(2005)] {slr} Schild, R.E.,
Leiter, D.J., \& Robertson, S.L., Observations Supporting the
Existence of an Intrinsic Magnetic Moment Inside the Central Compact
Object Within the Quasar Q0957+561, Astronomical J. (in press), 2005
(astro-ph/0505518)
\end{thebibliography}
 \end{document}